\newcommand{\qed}{\hbox{\rule[-2pt]{6pt}{6pt}}}
\documentstyle[eqsecnum,prd,aps,epsf]{revtex}
\draft
\begin{document}
\thispagestyle{empty}
{\baselineskip0pt
\leftline{\baselineskip16pt\sl\vbox to0pt{\hbox
{\it Department of Physics}
               \hbox{\it Waseda University}\vss}}
\rightline{\baselineskip16pt\rm\vbox to20pt{\hbox{WU-AP/143/02}
            \hbox{\today} 
\vss}}%
}
\vskip1cm
\begin{center}{\large \bf
No Go Theorem for Kinematic Self-Similarity with A Polytropic Equation of State}
\end{center}
\vskip1cm
\begin{center}
 {\large 
Hideki~Maeda$^a$
\footnote{Electronic address: hideki@gravity.phys.waseda.ac.jp}
, 
Tomohiro~Harada$^b$
\footnote{Electronic address: harada@gravity.phys.waseda.ac.jp}
, 
Hideo~Iguchi$^c$
\footnote{Electronic address: iguchi@th.phys.titech.ac.jp}
\\ and \\
Naoya~Okuyama$^d$
\footnote{Electronic address: okuyama@gravity.phys.waseda.ac.jp}\\
{\em ${}^{a,b,d}$Department of Physics,~Waseda University, Shinjuku, 
Tokyo 169-8555, Japan}\\
{\em ${}^{c}$Department of Physics, Tokyo Institute of Technology, \\
  Oh-Okayama, Meguro, Tokyo, 152-8550, Japan}}\\
\end{center}

\begin{abstract}
We have investigated spherically symmetric spacetimes which contain a perfect fluid obeying the polytropic equation of state and admit a kinematic self-similar vector of the second kind which is neither parallel nor orthogonal to the fluid flow. We have assumed two kinds of polytropic equations of state and shown in general relativity that such spacetimes must be vacuum, which is in contrast with in Newtonian gravity.
\end{abstract}


\section{Introduction}
There is no characteristic scale both in Newtonian gravity and in general relativity. A set of field equations is invariant under scale transformations if we assume appropriate matter fields. It implies the existence of scale-invariant solutions to the field equations. Such solutions are called self-similar solutions. Among them, a spherically symmetric self-similar system has been widely researched in the context both of Newtonian gravity and of general relativity. Although the self-similar solutions are only special solutions of the field equations, it has often been supposed that they play an important role in situations where gravity is an essential ingredient in a spherically symmetric system. In particular, {\it self-similarity hypothesis} has been proposed, which states that solutions may naturally evolve to self-similar form even if they start out more complicated in a variety of astrophysical and cosmological situations~\cite{carr1999}.

Self-similar solutions in Newtonian gravity have been researched in an effort to obtain realistic solutions of gravitational collapse leading to star formation~\cite{penston1969,larson1969,shu1977,hunter1977}. In the isothermal case ($p=c_s^2 \rho$, where $p$ is the pressure, $c_s$ is the sound speed and $\rho$ is the mass density), Larson and Penston independently found a self-similar solution, which is called the Larson-Penston solution, describing a gravitationally collapsing isothermal gas sphere~\cite{penston1969,larson1969}. Recent numerical simulations and results of mode analyses showed that the Larson-Penston solution is the best description for the central part of a collapsing gas sphere~\cite{ti1999,hn1997,hm2000a,hm2000b,fc1993,mh2001}. In the polytropic case ($p=K\rho^{\gamma}$, where $K$ and $\gamma$ are constants), Goldreich and Weber found the polytropic counterpart of the Larson-Penston solution describing a gravitationally collapsing polytropic gas sphere for the polytrope index $\gamma=4/3$ and showed that this solution is stable for spherical linear perturbations by means of a mode analysis~\cite{gw1980}. After that, other authors extended their work for general $\gamma$~\cite{yahil1983,ss1988,hm2000a}. We call these solutions the polytropic Larson-Penston solutions hereafter.

In general relativity, self-similarity is defined by the existence of a homothetic Killing vector field~\cite{CT1971}. Such self-similarity (or homothety) is called the first kind. Ori and Piran discovered the general relativistic counterpart of the Larson-Penston solution together with Hunter's family of solutions for a perfect fluid obeying an adiabatic equation of state $p=(\gamma-1)\mu$ $(1<\gamma\alt1.036)$, where $\mu$ is energy density~\cite{op1987,op1990}. They observed that a naked singularity forms in this solution for $1<\gamma \alt 1.0105$. Harada and Maeda found that generic non-self-similar spherical collapse converges to the general relativistic Larson-Penston solution in an approach to singularity for $1<\gamma\alt1.036$~\cite{hm2001,harada2001}. Since a naked singularity forms for $1<\gamma\alt1.0105$, this means the violation of cosmic censorship in spherically symmetric case (See also~\cite{harada1998}). This will be the strongest known counterexample against the cosmic censorship ever. This also provides strong evidence for the self-similarity hypothesis in general relativistic gravitational collapse. Then a natural question arises about whether collapsing self-similar solutions with a polytropic equation of state exist in general relativity. If such solutions do exist, they may play an important role in the final stage of generic collapse as the adiabatic case. 

In Newtonian gravity, self-similarity in the polytropic case has the different form of the dimensionless variable from that in the isothermal case since the sound speed is not constant in the polytropic case. The dimensionless variable is $c_s t/r$ in the isothermal case, while it is $[\sqrt{K}t^{2-\gamma}]/[G^{(\gamma-1)/2}r]$ in the polytropic case. The scaling functions of physical quantities in the latter case are of different forms from those in the former case~\cite{ss1988}. In general relativity, there exists a natural generalization of homothety called {\it kinematic self-similarity}, which is defined by the existence of a kinematic self-similar vector field~\cite{ch}. The natural general relativistic counterpart of the self-similarity in the Newtonian polytropic case is the kinematic self-similarity of the second kind in which the kinematic self-similar vector is `tilted', i.e. neither parallel nor orthogonal to the fluid flow. In the pioneering work by Benoit and Coley, they have investigated the spherically symmetric spacetimes which admit a `tilted' kinematic self-similar vector field of the second kind and contain a perfect fluid without restricting the specific form of the equation of state~\cite{bc1998} (See also~\cite{coley1997}).

In this paper, we study the spacetimes which admit a `tilted' kinematic self-similar vector field of the second kind and contain a perfect fluid obeying the polytropic equation of state. We assume two kinds of relativistic polytropic equations of state and show that there are no non-trivial kinematic self-similar solutions of the second kind obeying either of these polytropic equations of state. In this paper, we adopt the unit in which $c=1$.
      
\section{Spherically Symmetric Spacetime and Kinematic Self-Similarity}
The line element in a spherically symmetric spacetime is given by 

\begin{eqnarray}
ds^2 &=& -e^{2\Phi(t,r)}dt^2+e^{2\Psi(t,r)}dr^2+R(t,r)^2 d\Omega^2,
\end{eqnarray}
where $d\Omega^2=d\theta^2+\sin^2 \theta d\varphi^2$. We consider a perfect fluid as a matter field

\begin{eqnarray}
T_{\mu\nu} &=& p(t,r)g_{\mu\nu}+[\mu(t,r)+p(t,r)]U_{\mu} U_{\nu}, \\
U_{\mu} &=& (-e^{\Phi},0,0,0),
\end{eqnarray}
where $U^{\mu}$ is the four-velocity of the fluid element. We have adopted the comoving coordinates. In this paper, we assume two kinds of polytropic equations of state. One is 

\begin{eqnarray}
p=K \mu^{\gamma}, \label{eos1}
\end{eqnarray}
where $K$ and $\gamma$ are constants, and the other is~\cite{st}

\begin{eqnarray}
\left\{
\begin{array}{ll}
\displaystyle{p=K n^{\gamma}},\\
\displaystyle{\mu=m_b n+\frac{p}{\gamma-1}}.\label{eos2}
\end{array}
\right.
\end{eqnarray}
where the constant $m_b$ is the mean baryon mass and $n(t,r)$ is the baryon number density. We call the former equation of state (I) (EOS (I)) and the latter equation of state (II) (EOS (II)). Here we assume that $K \ne 0$ and $\gamma \ne 0,1$. We note the property of these equations of state. For $\gamma<0$, the fluid suffers from thermodynamical instability. For $0<\gamma<1$, both EOS's (I) and (II) are approximated by dust in high-density region, since $p/\mu=K\mu^{\gamma-1} \to 0$ for $\mu \to \infty$ for EOS (I) and 

\begin{eqnarray}
\frac{p}{\mu}&=&\frac{Kn^{\gamma-1}}{m_b+\frac{K n^{\gamma-1}}{\gamma-1}} \to 0 \quad \mbox{for $n \to \infty$},
\end{eqnarray}
for EOS (II). When $1<\gamma$, EOS (II) is approximated by $p=(\gamma-1)\mu$ in high-density region since 

\begin{eqnarray}
\mu=m_b n+\frac{K n^{\gamma}}{\gamma-1} \to \frac{K n^{\gamma}}{\gamma-1}=\frac{p}{\gamma-1} \quad \mbox{for $n \to \infty$}.
\end{eqnarray}
In the case of $2<\gamma$ for EOS (II) and $1<\gamma$ for EOS (I), the dominant energy condition can be violated in high-density region, which will be unrealistic. 

A kinematic self-similar spacetime is defined as a spacetime which admits a vector ${\bf\xi}$ such that 

\begin{eqnarray}
{\cal{L}}_{\bf\xi} h_{\mu\nu} =2 h_{\mu\nu},\quad
{\cal{L}}_{\bf\xi} U_{\mu} =\alpha U_{\mu},\label{kss}
\end{eqnarray}
where $h_{\mu\nu} =g_{\mu\nu}+U_{\mu}U_{\nu}$ is the projection tensor, ${\cal{L}}_{\bf\xi}$ denotes the Lie differentiation along ${\bf\xi}$ and $\alpha$ is a constant~\cite{ch,coley1997}. ${\bf\xi}$ is called a kinematic self-similarity vector field. We assume that the spacetimes admit a kinematic self-similar vector field. The similarity transformation is characterized by $\alpha$ which is referred to the similarity index. Here we treat only the case in which $\alpha \ne 0, 1$, corresponding to self-similarity of the second kind, and a kinematic self-similar vector is `tilted'. Other cases are treated in~\cite{mhio2002}. In the spherically symmetric case, the `tilted' kinematic self-similar vector ${\bf\xi}$ of the second kind can be written as

\begin{eqnarray}
\xi^{\mu}\frac{\partial}{\partial \xi^{\mu}}=\alpha t\frac{\partial}{\partial t}+r \frac{\partial}{\partial r},
\end{eqnarray}
and self-similarity implies that the metric functions are given by

\begin{eqnarray}
R=r S(\xi), \quad \Phi=\Phi(\xi), \quad \Psi=\Psi(\xi),
\end{eqnarray}
where $\xi=r/(\alpha t)^{1/\alpha}$ is the self-similar variable. The Einstein equations imply that the quantities $m, \mu$ and $p$ must be of the form 

\begin{eqnarray}
2Gm&=&r\left[M_1(\xi)+\frac{r^2}{t^2}M_2(\xi)\right],\label{2ndm}\\
8\pi G \mu&=&\frac{1}{r^2}\left[W_1(\xi)+\frac{r^2}{t^2}W_2(\xi)\right],\label{2ndmu}\\
8\pi G p&=&\frac{1}{r^2}\left[P_1(\xi)+\frac{r^2}{t^2}P_2(\xi)\right].\label{2ndp}
\end{eqnarray}
These functions are the same as those defined by Benoit and Coley in~\cite{bc1998}. A set of ordinary differential equations can be obtained by demanding that the Einstein equations and the equations of motion of the matter field are satisfied for the $O[(r/t)^0]$ and the $O[(r/t)^2]$ terms separately. Then the Einstein equations and the equations of motion for the perfect fluid are reduced to the equations (A.7)-(A.17) in~\cite{bc1998}. For convenience, we also use the $(tt)$ and $(rr)$ components of the Einstein equations which reduce to the following:

\begin{eqnarray}
S'(S'+2\Psi'S)&=&\alpha^2W_2 S^2 e^{2\Phi}, \label{2nd00a}\\
2S(S''+2S')-2\Psi'S(S+S')&=&-S'^2-S^2+e^{2\Psi}(1-W_1S^2),\label{2nd00b}\\
2S(S''+\alpha S'-\Phi'S')+S'^2&=&-\alpha^2 P_2S^2 e^{2\Phi},\label{2nd11a}\\
(S+S')(S+S'+2\Phi'S)&=&(1+P_1S^2)e^{2\Psi},\label{2nd11b}
\end{eqnarray}
where the prime denotes the derivative with respect to $\ln \xi$. These equations are dependent on equations (A.7)-(A.17) in~\cite{bc1998}

In a vacuum case, while the Minkowski spacetime can be obtained for all $\alpha$, the Schwarzschild spacetime can be obtained only for $\alpha=3/2$ since equations (A.8) and (A.10) in~\cite{bc1998} are degenerated so that $M_2 \ne 0$ is possible in this case. The Schwarzschild spacetime in the Lemaitre's choice of coordinates is written as

\begin{eqnarray}
ds^2=-dt^2+r_g^{\frac23}\left(\frac{dR^2}{\left[\frac32(R-t)\right]^{\frac23}}+\left[\frac32(R-t)\right]^{\frac43}d\Omega^2\right),
\end{eqnarray}
with $r_g=(3/2)(R-t)$ being the Schwarzschild radius~\cite{LL}. Changing the radial coordinate as $R=r^{3/2}$, 

\begin{eqnarray}
ds^2=-dt^2+r_g^{\frac23}\left(\frac{(9/4)dr^2}{\left[\frac32(1-t/r^{\frac32})\right]^{\frac23}}+r^2\left[\frac32(1-t/r^{\frac32})\right]^{\frac43}d\Omega^2\right),
\end{eqnarray}
can be obtained. 

\section{No Go Theorem}

{\bf Theorem}.\\
{\it Let the (${\cal M}^4$,g) be a spherically symmetric spacetime which\\
\quad {\rm a)} admits a kinematic self-similar vector of the second kind $\xi^{\mu}$ with similarity index $\alpha \ne 0,1$ which is neither parallel nor orthogonal to the fluid flow; and\\
\quad {\rm b)} satisfies the Einstein equations for a perfect fluid obeying equation of state {\rm(I)} or {\rm(II)} for $K \ne 0$ and $\gamma \ne 0, 1$.\\
Then (${\cal M}^4$,g) is the Minkowski spacetime with any $\alpha$ or the Schwarzschild spacetime with $\alpha=3/2$.}\\
\\
{\it Proof}. Subtracting equation (\ref{2nd11b}) from equation (\ref{2nd00b}) and eliminating $S''$ with using equation (A.17) in~\cite{bc1998}, we obtain

\begin{eqnarray}
2\Phi'&=&(P_1+W_1)e^{2\Psi}. \label{2ndkey1}
\end{eqnarray}
Then equations (A.13) and (A.14) in~\cite{bc1998} result in

\begin{eqnarray}
e^{2\Psi}(P_1+W_1)^2&=&4P_1-2P_1',\label{2ndkey2}\\
e^{2\Psi}(P_1+W_1)(P_2+W_2)&=&-2P_2'.\label{2ndkey3}
\end{eqnarray}

If a perfect fluid obeys EOS (I) for $K \ne 0$ and $\gamma \ne 0, 1$, we find from equations (\ref{2ndmu}) and (\ref{2ndp}) that

\begin{eqnarray}
\alpha=\gamma, \quad P_1=W_2=0,\quad P_2 =\frac{K}{(8\pi G)^{\gamma-1}\gamma^2}{\xi}^{-2\gamma}W_1^{\gamma} \quad \mbox{[case (A)]},\label{casea}
\end{eqnarray}
or
\begin{eqnarray}
\alpha=\frac{1}{\gamma},\quad P_2=W_1=0,\quad P_1 =\frac{K}{(8\pi G)^{\gamma-1}\gamma^{2\gamma}}{\xi}^2 W_2^{\gamma} \quad \mbox{[case (B)]}.\label{caseb}
\end{eqnarray}
If a perfect fluid obeys EOS (II) for $K \ne 0$ and $\gamma \ne 0, 1$, we find from equations (\ref{2ndmu}) and (\ref{2ndp}) that

\begin{eqnarray}
\alpha=\gamma,\quad P_1=0,\quad P_2 =\frac{K}{m_b^{\gamma}(8\pi G)^{\gamma-1}\gamma^2}{\xi}^{-2\gamma}W_1^{\gamma}=(\gamma-1)W_2 \quad \mbox{[case (C)]},\label{casec}
\end{eqnarray}
or
\begin{eqnarray}
\alpha=\frac{1}{\gamma},\quad P_2=0,\quad P_1 =\frac{K}{m_b^{\gamma}(8\pi G)^{\gamma-1}\gamma^{2\gamma}}{\xi}^2 W_2^{\gamma}=(\gamma-1)W_1 \quad \mbox{[case (D)]}.\label{cased}
\end{eqnarray}

$P_1=0$ directly implies $W_1=0$ from equation (\ref{2ndkey2}), while $P_2=0$ implies $(P_1+W_1)W_2=0$ from (\ref{2ndkey3}), which results in $W_2=0$ for cases (B) and (D). Therefore the spacetime must be vacuum for all possible cases. $\qed$

\section{Summary and Discussion}
We have shown that there are no spherically symmetric solutions that contain a perfect fluid obeying either polytropic EOS (I) or (II) and admit a `tilted' kinematic self-similar vector of the second kind. There exist only vacuum solutions, the Minkowski solution with any $\alpha \ne 0,1$ and the Schwarzschild solution with $\alpha=3/2$. The result is much different from that in Newtonian case, since in Newtonian case there exists the polytropic Larson-Penston solution, which is a self-similar solution describing the collapsing polytrope gas~\cite{gw1980,yahil1983,ss1988,hm2000a}. 

The result in this paper does not directly mean that the kinematic self-similarity of the second kind is incompatible with perfect fluids since the realistic equation of state in relativistic regime could be neither EOS (I) nor (II). The proof of the no go theorem strongly depends on the fact that $P_1=0$ or $P_2=0$ and the theorem has not been proved in more general cases. The non-trivial solutions could be obtained when we assume the equation of state in which $P_1 P_2 \ne 0$ is possible. Such cases will be investigated elsewhere.

EOS's (I) and (II) could be compatible with other kinds of self-similarities, i.e. the zeroth and the infinite kinds, or with the case which the kinematic self-similar vector is parallel or orthogonal to the fluid flow. These cases are considered in~\cite{mhio2002}. 

If the polytropic Larson-Penston solution is an attractor of generic collapse of the polytropic gas in Newtonian gravity as in the isothermal case, how does the collapse proceeds in the relativistic region when we assume that the polytropic equation of state in relativistic regime is described as EOS (I) or (II)? For $0<\gamma<1$, both EOS's (I) and (II) are approximated by dust so that the generic collapse could converge to the spacetime of which the central region can be described by the Tolman-Bondi solution~\cite{LTB}.

In order to show the whole picture of the generic collapse of the polytrope gas which obeys equation of state (I) or (II), the full numerical simulations of gravitational collapse will be quite helpful. 

\acknowledgments

We are very grateful to T.~Torii for helpful comments. We would also like to thank K.~Maeda for continuous encouragement. 
This work was partly supported by the 
Grant-in-Aid for Scientific Research (Nos. 05540 and 11217)
from the Japanese Ministry of
Education, Culture, Sports, Science and Technology.

\end{document}